\documentclass{mybeamer2}

\usepackage{epic}
\usepackage{eepic}
\usepackage{color}

\def\a{\alpha}

\def\o{\omega}

\def\e{\varepsilon}

\def\be{\begin{equation}}
\def\ba{\begin{array}}
\def\ee{\end{equation}}
\def\ea{\end{array}}
\def\bea {\begin{eqnarray}}
\def\eea {\end{eqnarray}}
\def\bean{\begin{eqnarray*}}
\def\eean{\end{eqnarray*}}

\def\RA {\ \Rightarrow\ }

\title{A toy model of wave turbulence}

\author{Elena Kartashova
  \\ \footnotesize{Institute for Analysis, J. Kepler University, Linz, Austria\\
  Elena.Kartaschova@jku.at}
  }
\institute{April 12, 2011 \\
KITP, Santa Barbara}

\begin{document}

\begin{frame}{}
\titlepage
\end{frame}

\section{Classical  wave turbulence}
\begin{frame}{}
\begin{center}
 Kolmogorov-Zakharov energy spectra
\end{center}
\begin{columns}

\column{5.0cm}
\textbf{ASSUMPTIONS}
\begin{itemize}
\item weak nonlinearity, $0< \e \ll 1$,
\item randomness of phases,
\item infinite-box limit, $L/\lambda\to \infty,$
\item existence of inertial interval $(k_1,k_2)$,
\item locality in $k$-space (waves with wavelengths of the same order $k$ do interact),
\item interactions are locally isotropic (no dependence on direction)
\end{itemize}
\column{6.5cm}
\begin{figure}
\vskip -1.2cm
\begin{center}
\includegraphics[width=5cm,height=3cm]{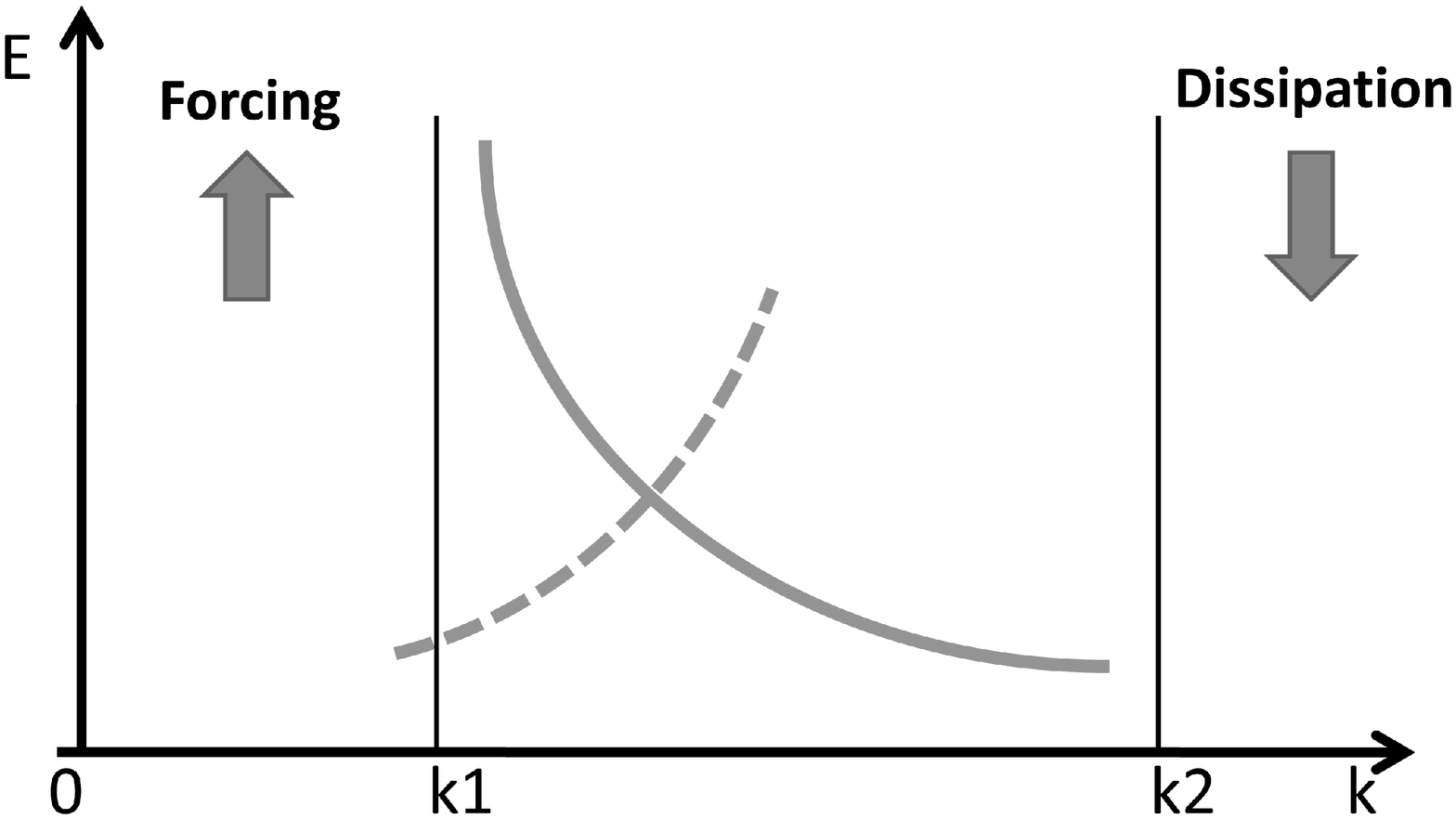}
\end{center}
\end{figure}

\textbf{ADVANTAGES}
\vskip 0.5cm
\begin{itemize}
\item SIMPLE FORMULA: $E \sim k^{-\nu}, \nu>0$, $\nu$ is \alert{constant for a given wave system}.
\item KZ-spectra \alert{do not depend on the initial conditions}.
\end{itemize}
\end{columns}
\end{frame}

\begin{frame}{}
\begin{center}
THE BAD NEWS: THERE ARE PROBLEMS
\end{center}
Results of laboratory experiments:

\begin{itemize}
\item{}\textbf{energy cascade}
\begin{itemize}
\item{} \textbf{no cascade}: Hammack \emph{et al.}.
Fluid Mech., 2005), regular patterns, {\color{green}surface water waves};
\item{} \textbf{cascade consists of two parts: discrete and continuous}: {\color{red}great amount of experiments in various wave systems};
\item{} \textbf{continuous part of energy spectrum is not KZ-spectrum}: Mordant, PRL, 2008; {\color{yellow}a thin elastic steel plate} is excited with a vibrator; Falcon \emph{et al.}, PRL, 2007 ({\color{blue}grav., grav.-cap., cap., mercury})
\end{itemize}
\item{}\textbf{wave interactions are not local}: Abdurakhimov \emph{et al}, J. Phys.: Conf. Ser., 2009;  {\color{cyan}capillary waves in He-II}.
\item{}\textbf{form of energy spectra depends on initial conditions}: Falcon \emph{et al.}, PRL, 2007 ({\color{blue}grav., grav.-cap., cap., mercury}); Xia \emph{et al.}, EPL, 2010 ({\color{gray}capillary water waves}).

\begin{center}
\alert{\textbf{Something is rotten in the state of Zakharov!}}
\end{center}
\end{itemize}
\end{frame}

\begin{frame}{}
\begin{center}
ATTEMPTS TO SOLVE THE PROBLEMS:
\end{center}
\begin{itemize}
\item{} {\color{green} frozen  WT} (Pushkarev+Zakharov, Physica D, 2000)
\item{} {\color{blue} sandpile model of WT} (Nazarenko, J. Stat. Mech.: Theor. Exp., 2006)
\item{} {\color{yellow} mesoscopic WT} (Zakharov \emph{et al.}, JETP Lett., 2005)
\item{} {\color{magenta} laminated WT}(K, JETP Lett., 2006)
\begin{itemize}
\item{} discrete layer - {\color{cyan} discrete WT} (K, PRL 2007; EPL 2009; K+L'vov, PRL 2007; K, Cambridge University Press, 2010)
\item{} continuous layer - {\color{gray} classical WT} (Zakharov, L'vov, Falkovich, Springer, 1992)
\end{itemize}
\item{} {\color{purple} finite-dimensional WT} (L'vov \emph{et al.}, PRE, 2009)
\end{itemize}

However, no model gives \alert{a general answer} to a simple question:
\begin{itemize}
\item{} \textbf{\emph{how to describe time evolution of a wave system beginning with one initially excited wave?}}
\end{itemize}

\alert{A partial answer} is given by the model of discrete wave turbulence - in terms of resonance clusters.
\end{frame}
\section{Discrete wave turbulence (DWT)}
\begin{frame}{DWT - brief overview}
\begin{itemize}
\item{} 1. solve resonance conditions in \alert{\textbf{integers}} (a tricky thing due to \emph{Hilbert's 10th Problem})
\item{} 2. construct an NR-diagram for each resonance cluster;
 \item{} 3. write out explicit form of dynamical systems for each resonance cluster (automatically follows from the form of an NR-diagram)
\end{itemize}
\begin{figure}
\begin{center}
\includegraphics[width=5cm,height=3cm]{KZ-classical.eps},
\includegraphics[width=5cm,height=3cm]{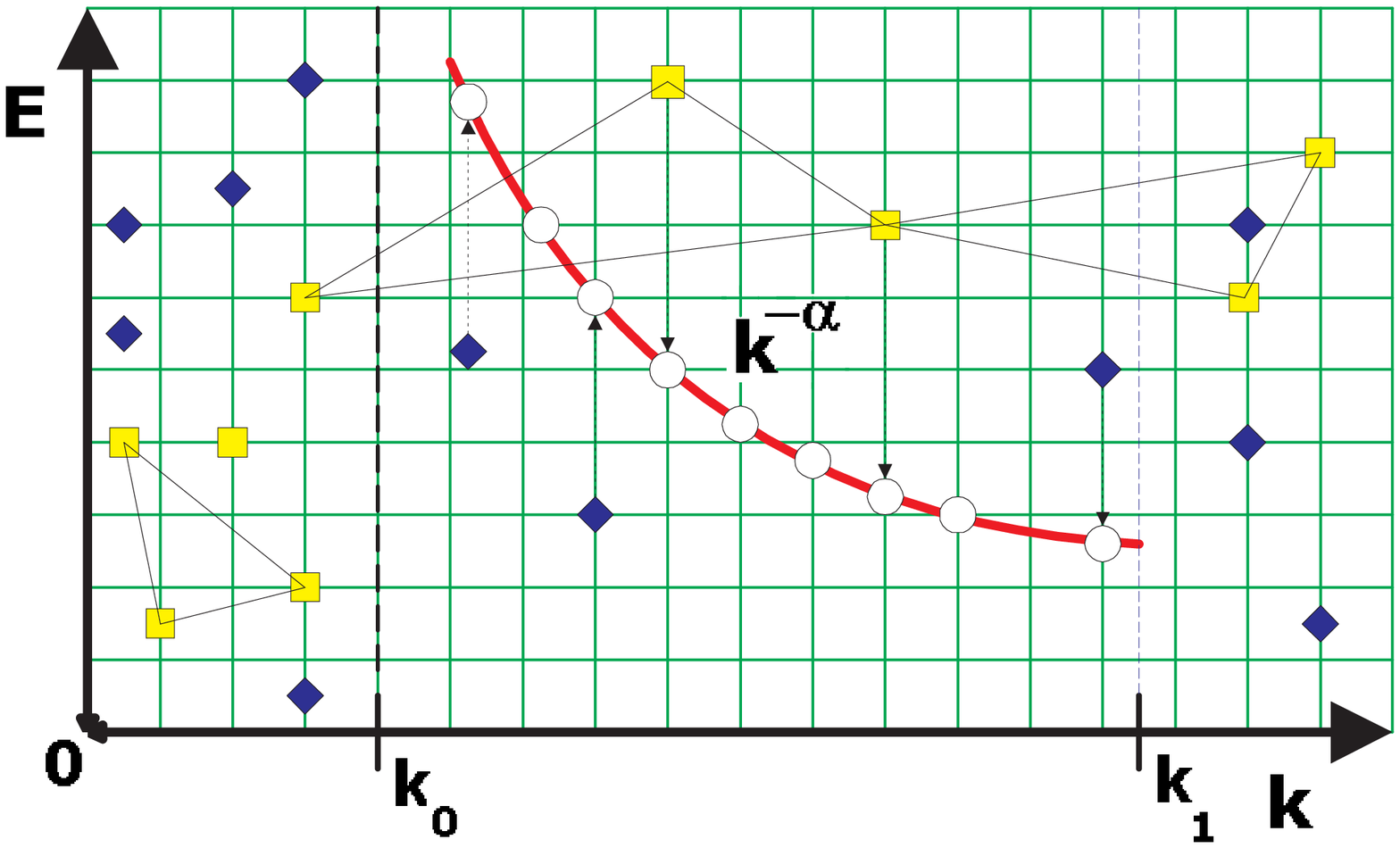}
\end{center}
\end{figure}

Kartashova, Nonlinear resonance Analysis: Theory, Computation, Applications (Cambridge University Press, 2010)

\end{frame}

\begin{frame}{}
\begin{center}
\textbf{1. Solution of resonance conditions, the idea}
\end{center}
Take 2D surface water waves, $\o \sim \sqrt[4]{m^2+n^2}$, frequency res. condition
\be \label{tralala}\o_1+\o_2=\o_3+\o_4 \ee
\emph{\textbf{Brute-force computation}}: \alert{3 days} for $m,n \le 128$.

\textbf{\emph{q-class decomposition}}: \alert{3 minutes} for $m,n \le 1000$.

The idea of q-class decomposition:
\be a\sqrt{3}+b\sqrt{5}=0 \ee
has no solutions with integer $a$ and $b$.

Regard presentation (\textbf{\emph{it is unique!}}):
\be
\sqrt[4]{m^2+n^2}=\gamma \sqrt[4]{q}, \quad q=q_1^{\a_1}\cdots q_n^{\a_n}, \ \a_j \le 3,
\ee
then (\ref{tralala}) has solutions (\emph{\textbf{necessary condition!}}) only if 1) all 4 wavevectors have the same $q$, or 2) they have pairwise equal $q$-s, i.e. $q_1=q_3$ and $q_2=q_4$.

Generalization for \alert{arbitrary finite number of different radicals} -  Besicovitch theorem (J. Lond. Math. Soc., 1940)

\end{frame}

\begin{frame}{}
\begin{center}
\textbf{2. Structure of resonances}
\end{center}

\begin{columns}
\column{6cm}
\vskip -0.5cm
\textbf{Geometrical structure}
\begin{figure}
\includegraphics[width=4.5cm,height=2.5cm]{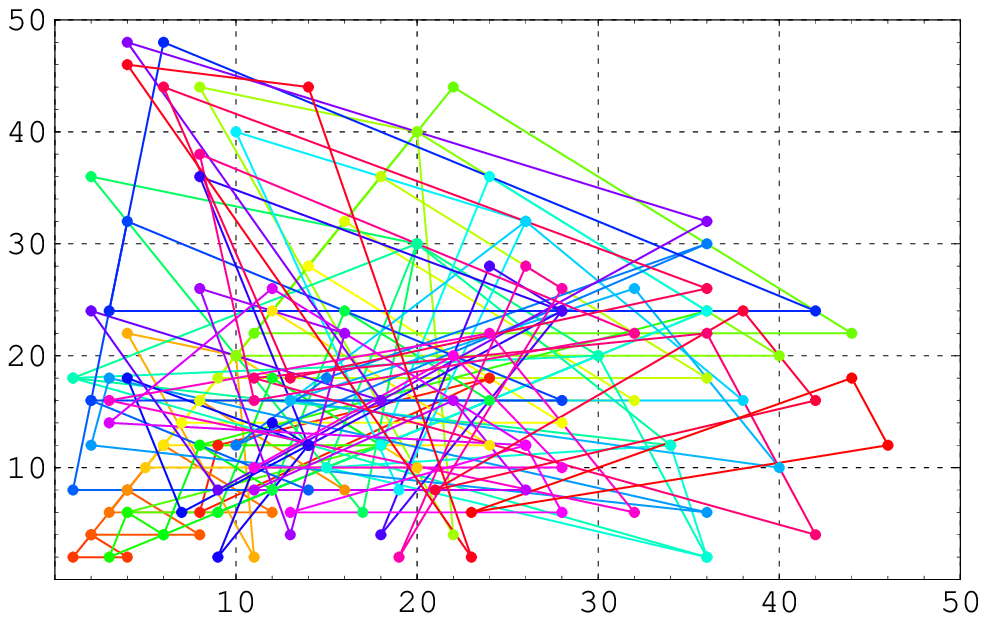}
\end{figure}
\column{6cm}
\vskip -0.5cm
\textbf{Topological structure}
\begin{figure}
\vskip -0.15cm
\includegraphics[width=4.5cm,height=2.3cm]{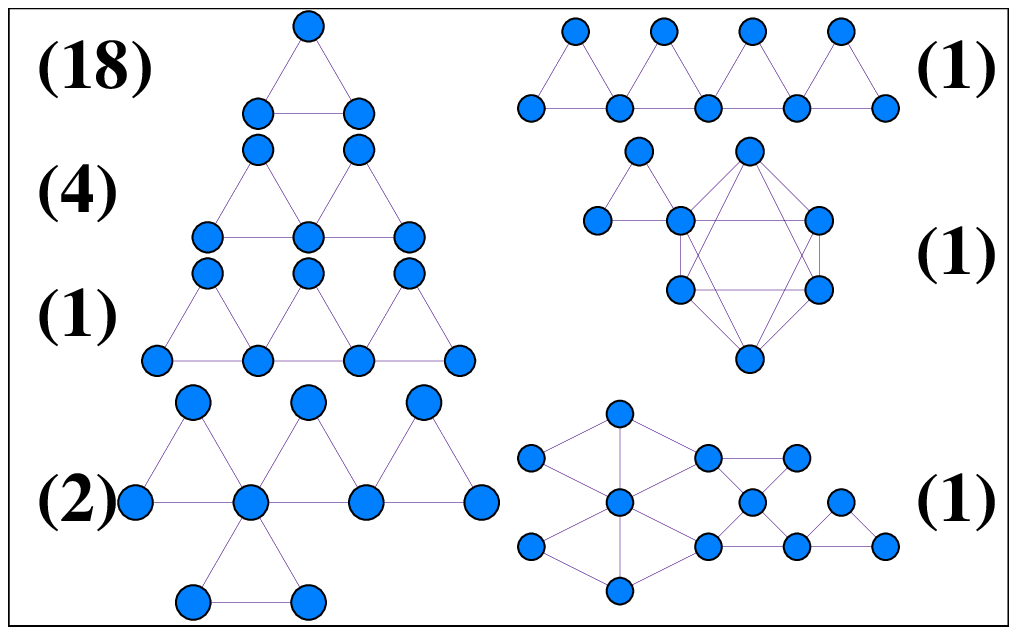}
\end{figure}
\end{columns}
\begin{itemize}
\item Altogether 2500 Fourier modes in spectral domain $m,n \le 50$ for $\o\sim 1/\sqrt{m^2+n^2}$
\item Only 128 take part in resonances - $\sim 5\% $ of all modes
\item 28 clusters - 18 are \textbf{\alert{integrable!}}, $\sim 60\%; $ max cluster - 12 modes
\end{itemize}
\begin{figure}
\vskip -0.7cm
\includegraphics[width=2.5cm,height=4cm,angle=270]{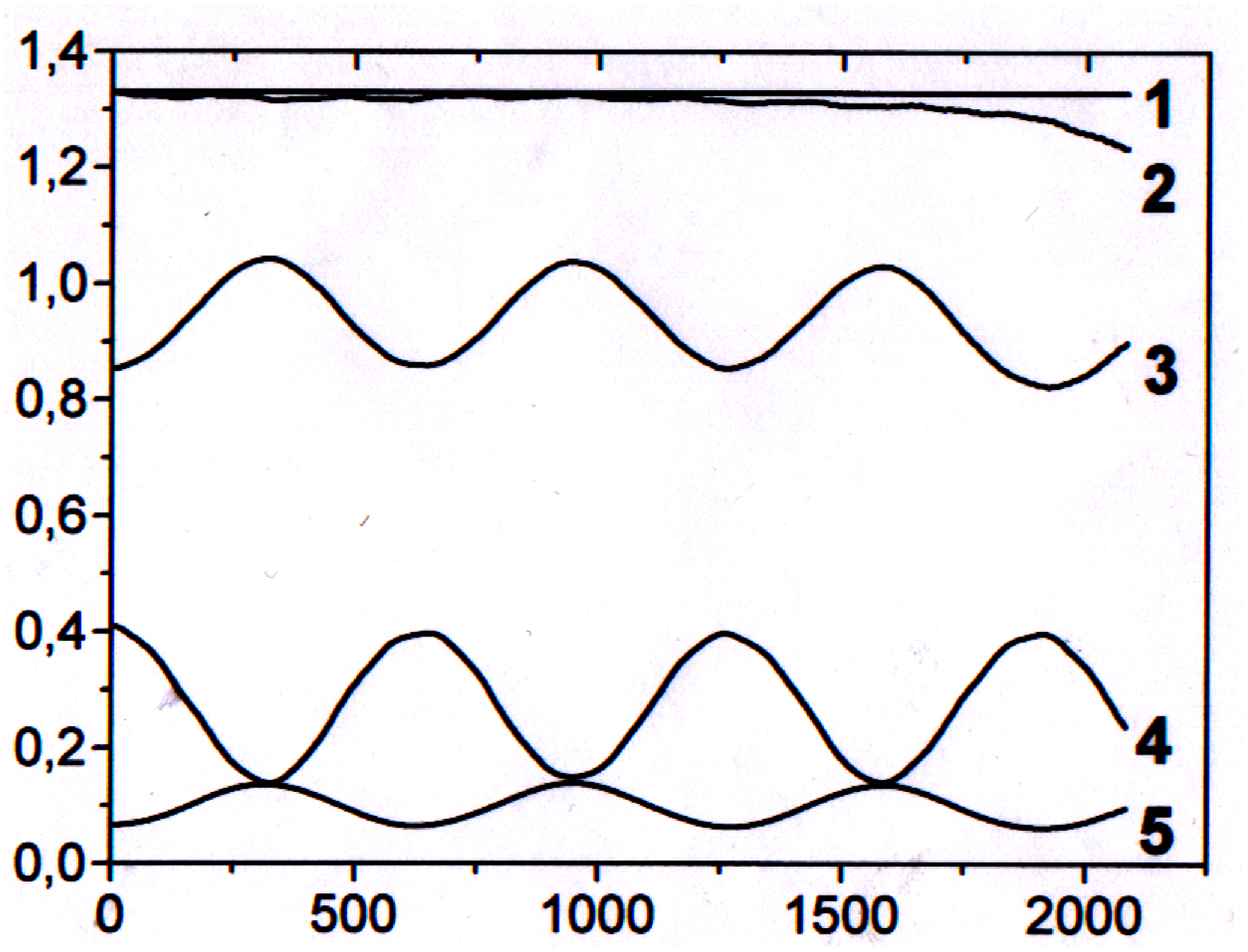}
\includegraphics[width=2.5cm,height=4cm,angle=270]{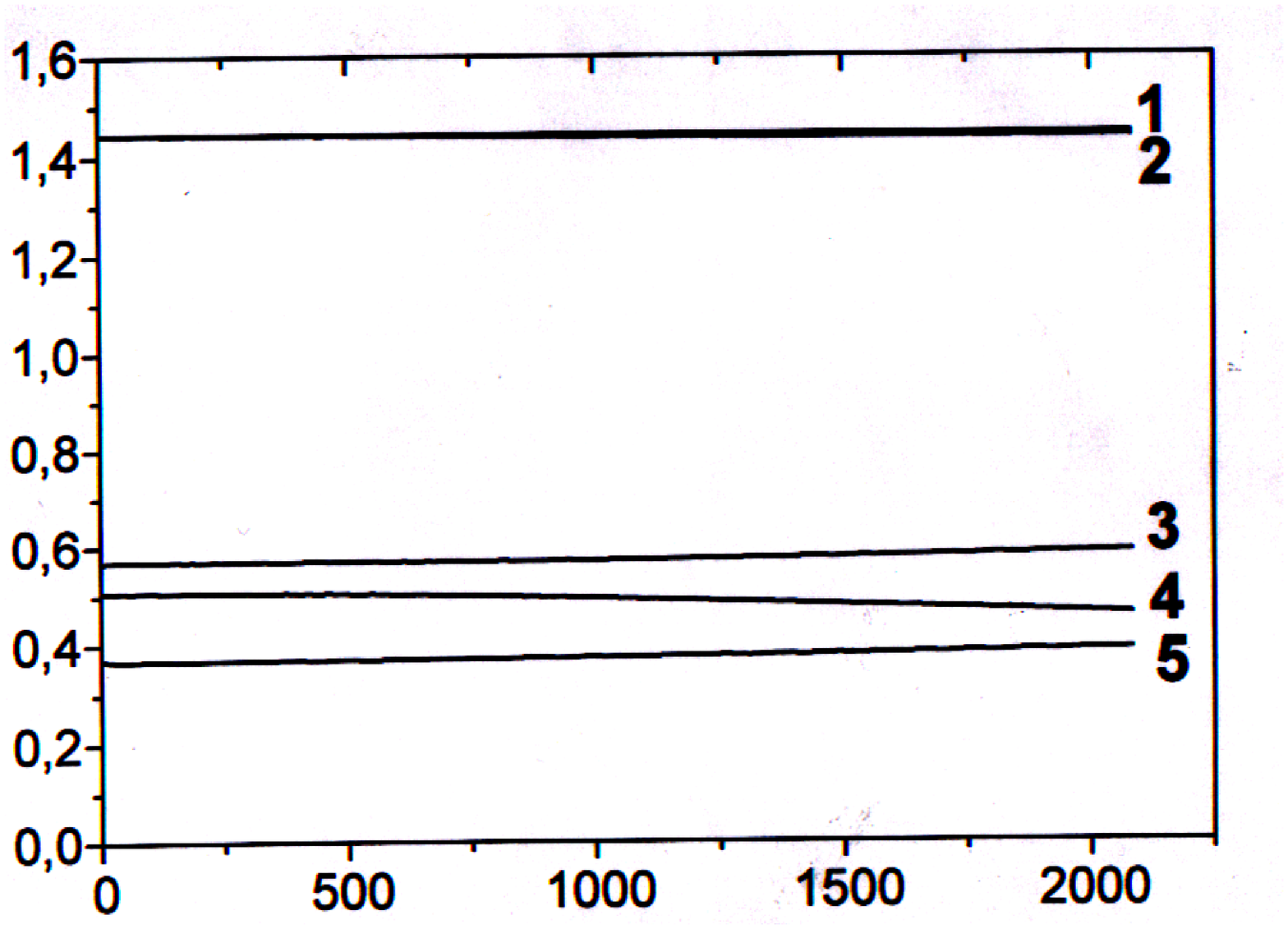}
\end{figure}
Kartashova (PRL, 1994)
\end{frame}

\begin{frame}{}
\begin{center}
\textbf{NR-diagram and dynamical system}
\end{center}

\begin{columns}
\column{6.5cm}
\alert{\emph{A}}-mode -- bold, \emph{P}-mode -- dashed:
\column{5cm}
\begin{figure}
\vskip -1.2cm
\begin{center}
\includegraphics[width=4.cm,height=1.3cm]{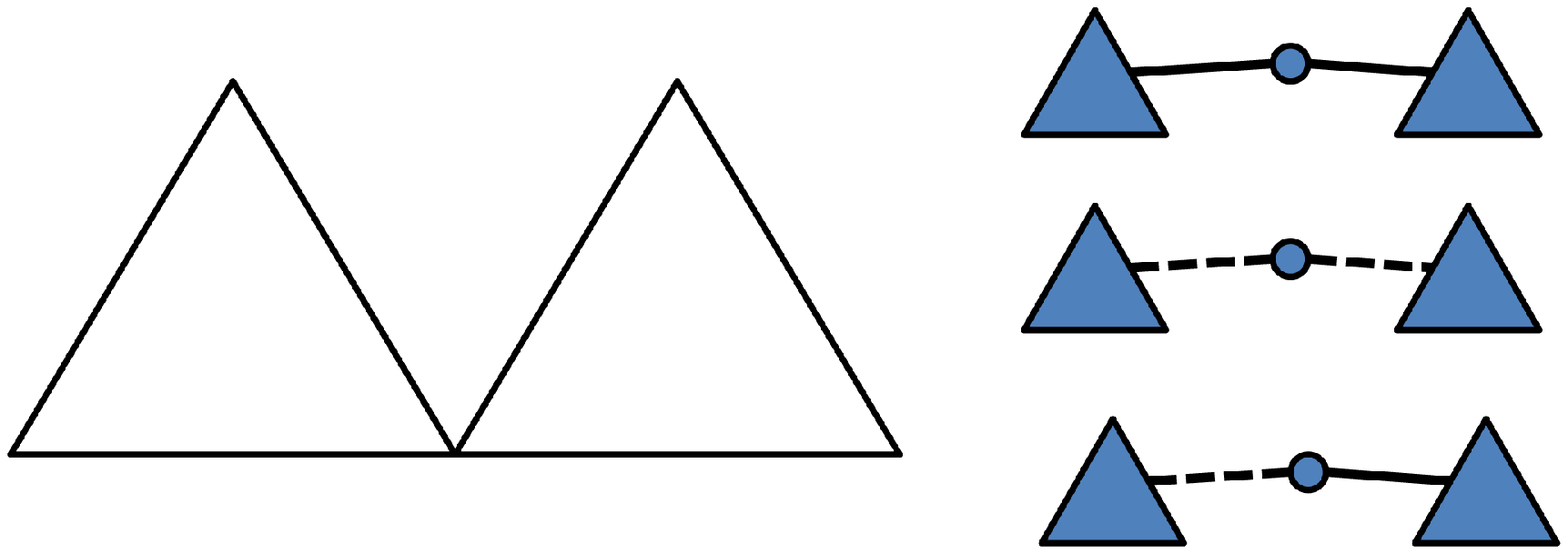}
\end{center}
\end{figure}
\end{columns}
\begin{figure}
\vskip -0.2cm
\begin{center}
\begin{tabular}{|c|c|c|c|}
  \hline
  ~~\includegraphics[width= 2.5cm,height=1.7cm]{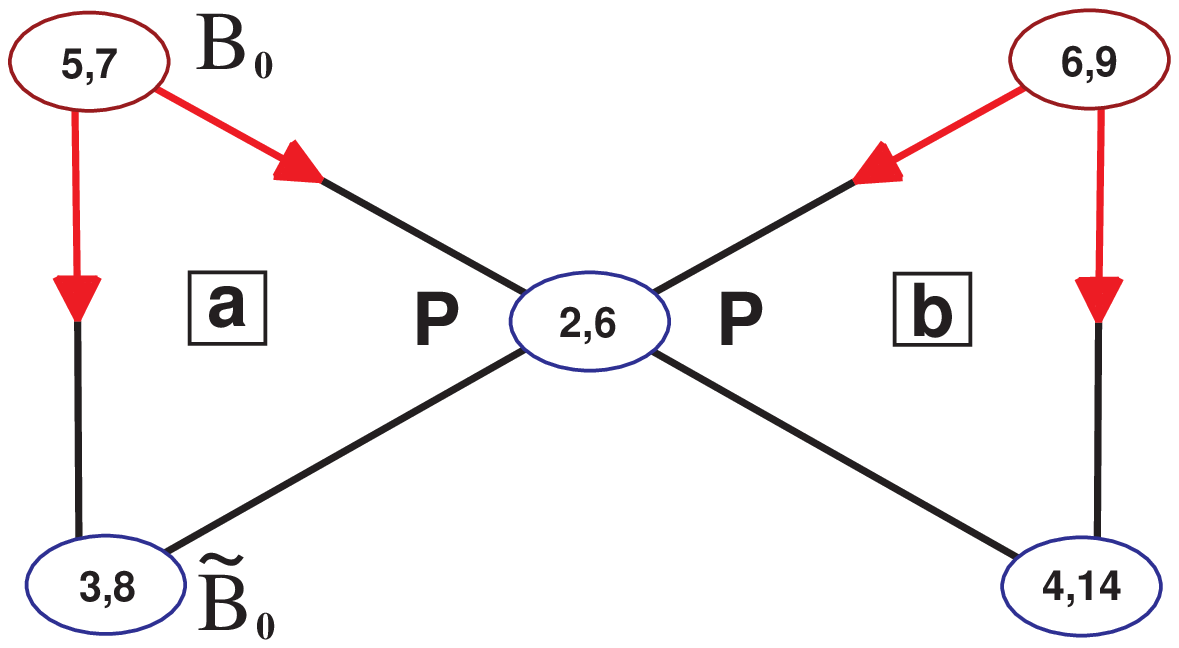}~~ &  
  ~~\includegraphics[width= 2.5cm,height=1.7cm]{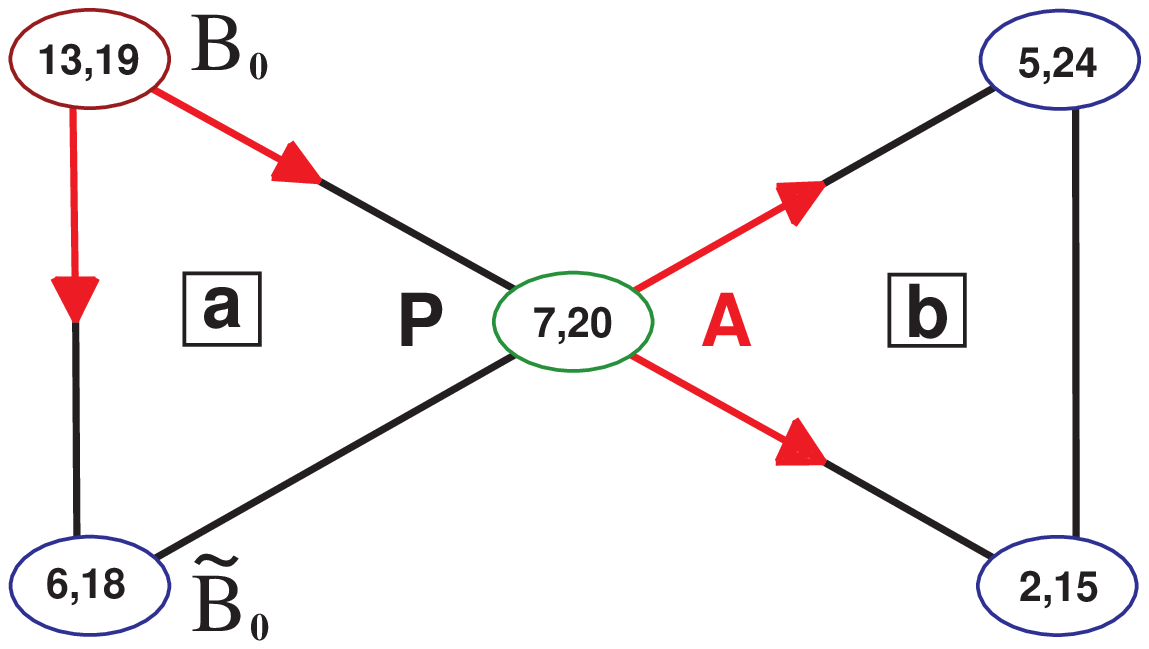}~~&   
 ~~\includegraphics[width= 2.5cm,height=1.7cm]{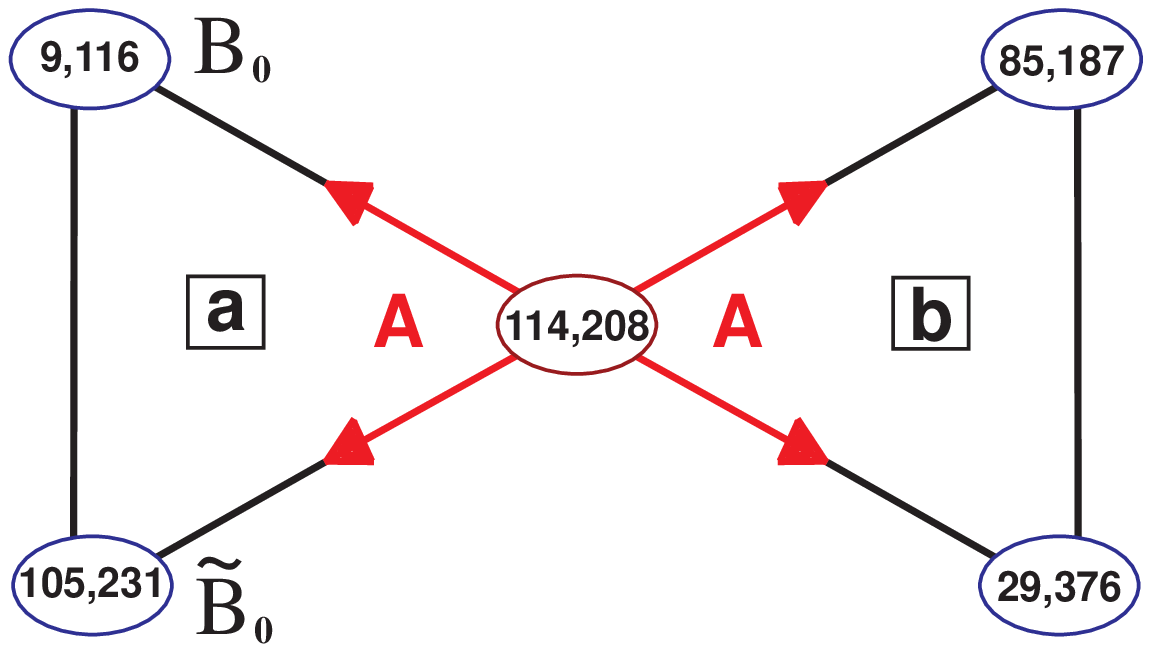}~ \\   
PP-butterfly   & PA-butterfly &  AA-butterfly \\  \hline
\end{tabular}
\end{center}
\end{figure}
NR-diagram defines \emph{\textbf{uniquely}} the form of dynamic system of resonance cluster and conservation laws, e.g. for PA-butterfly:
\bea \label{2}
\dot{B}_{1|b}&=&  Z_b B_{2|b}^*B_{3|b}\,, \  \dot{B}_{3|a}=  - Z_a B_{3|b} B_{2|a}\,, \
  \dot{B}_{2|b}=  Z_b B_{1|b}^* B_{3|b}\,,\\   \dot{B}_{2|a}&=&  Z_a B_{3|b}^* B_{3|a}\,,\
\dot{B}_{3|b}=  - Z_b B_{1|b} B_{2|b}  + Z_a B_{2|a}^* B_{3|a}\,\eea
and conservation laws read
\be
  I_{a}= |B_{2|a} |^2 + |B_{3|a}|^2 \,, \
  I_{b}=|B_{1|b} |^2 - |B_{2|b}|^2 \,, \
   I_{a,b}=|B_{1|b}|^2+|B_{3|b} |^2 + |B_{3|a}|^2\,. \nonumber
\ee

\end{frame}
\begin{frame}{}
\begin{center}
\textbf{Lab. experiments} (Chow, Henderson, Segur, Fluid Mech, 1996)
\end{center}
\emph{Five} frequencies  but \emph{seven} different modes (2\textbf{D} gravity-capillary waves). The amplitudes and frequencies were identified as
\bea A_1 \leftrightarrow 60 \mbox{ Hz}, \quad A_2 \leftrightarrow 35 \mbox{ Hz}, \quad A_3 \leftrightarrow \alert{25} \mbox{ Hz},\nonumber \\ \quad A_4 \leftrightarrow \alert{25} \mbox{ Hz}, \quad A_5 \leftrightarrow 10 \mbox{ Hz}, \quad A_6 \leftrightarrow \alert{25} \mbox{ Hz}, \quad A_7 \leftrightarrow 15 \mbox{ Hz}, \nonumber \eea
(\emph{modes with the same frequencies may have different wavevectors!}) and resonance conditions for frequencies as
\be
\o_1=\o_2+\o_3, \quad \o_2=\o_4+\o_5, \quad \o_6=\o_5+\o_7
\ee
with $\o_3=\o_4=\o_6.$
\begin{figure}[h]
\begin{center}
\includegraphics[width=3.cm,height=1.6cm]{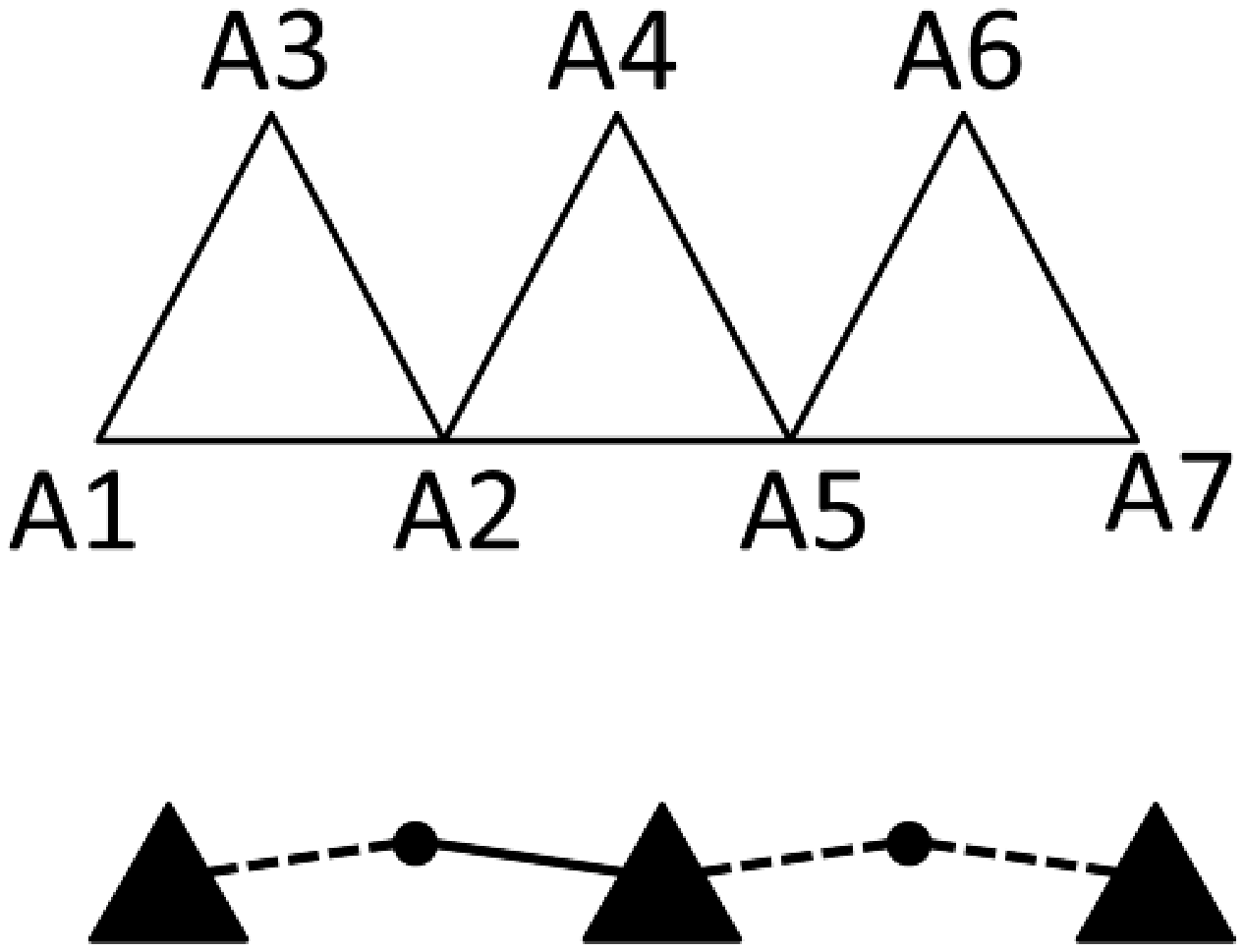}
\end{center}
\end{figure}
\alert{NR-diagram gives more information than simple frequency analysis!}
\end{frame}

\begin{frame}{}

What do we know now:
\begin{itemize}
\item{} \textbf{If initially excited mode is \alert{resonant or near-resonant}, DWT gives the answer about further evolution of a wave system in the form of some explicitly given dynamical systems, most of them integrable!}
\end{itemize}

What else do we need to know:
\begin{itemize}
\item{} \textbf{If initially excited mode is \alert{NOT} resonant or near-resonant - what happens?}
\end{itemize}
\end{frame}

\section{Energy cascades in DWT}

\begin{frame}{One-wave instability}
\textbf{} 

 Examples:
\begin{itemize}
\item parametric instability in classical mechanics,
\item  Suhl instability  of spin waves,
\item Oraevsky-Sagdeev  decay instability of plasma waves,
\item modulation instability in nonlinear optics,
\item Benjamin-Feir instability in deep water, etc.
\end{itemize}

It is described at early stages of the process as interaction of three monochromatic wave trains: carrier ($\o_c$ ), upper ($\o_+=\o_c + \Delta \o$) and lower ($\o_-=\o_c - \Delta \o$) side-band waves with small $\Delta \o >0$ which form a  quartet for one particular configuration which occurs when two of the waves coincide, with frequency resonance condition
\be \label{Phil}
\o_+ + \o_- = 2\o_c.
\ee

\end{frame}

\begin{frame}{}
\begin{center}
\textbf{Cascading cluster}
\end{center}
\bea \label{general}
\begin{cases}
\o_{f}= \boxed{\o_{1,1}}+\o_{2,1},  \, \quad \quad \quad \ \ \, E_1=p_1E_f, \, 0< p_j < 1, \\
\boxed{\o_{1,1}}=\o_{2,1}+\o_{2,2}, \,  \, \quad \quad \quad E_2=p_2E_1, \\
\o_{2,1}=\o_{3,1}+\o_{3,2},  \, \quad \ \ \quad \quad \ E_3=p_3E_2,\\
....\\
\o_{n-1,1}=\o_{n,1}+ \o_{n,2},  \, \quad \quad \quad \, E_n=p_n E_{n-1}
\end{cases}
\eea
\begin{figure}
 \vskip -0.5cm
\includegraphics[width=8cm,height=5cm]{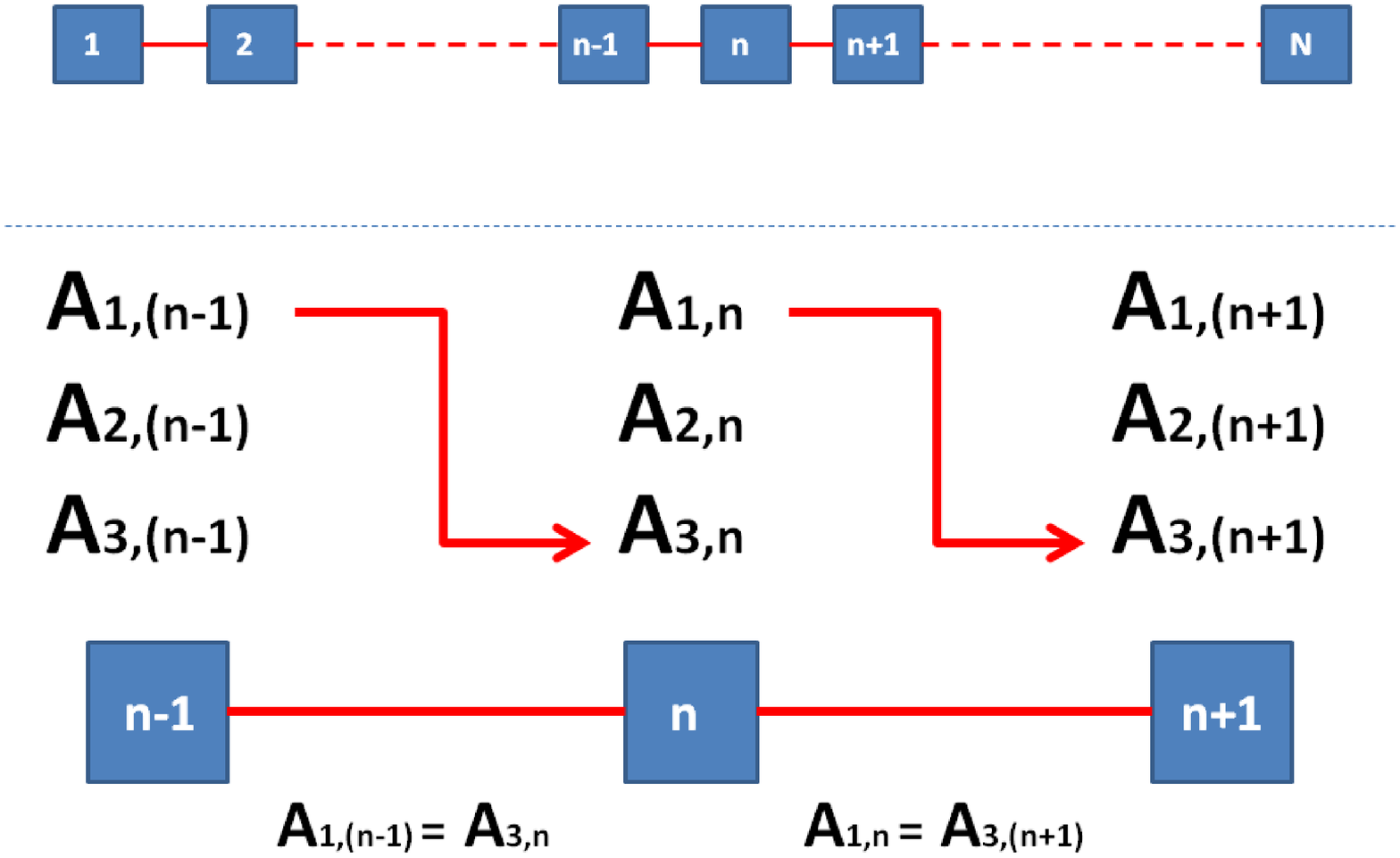}
\end{figure}
\end{frame}

\begin{frame}{}
\begin{center}
\textbf{\alert{First assumption: intensity of a cascade $p$ is constant}}
\end{center}
\be \label{p}
E_n =p^n E_0 \RA A_{n+1}=\sqrt{p} A_n
\ee
and total energy of one cascading chain reads
\be \label{E-chain}
E = \lim_{n \rightarrow \infty} \sum_{n} A_o^2 p^n = \frac{A_0^2}{1-p}=\mbox{const for} \quad p<1.
\ee
\end{frame}

\begin{frame}{}
\begin{center}
\textbf{\alert{Second assumption: each cascading mode is excited when corresponding increment of instability is maximal}}
\end{center}
Maximal increment at $n$-th cascade step can be  computed (BF-form) as
\be \label{BFI-incr-max}
I_{\mathbf{max},n}=\frac{|(\Delta \o)_n|}{\o_n A_n k_n}= 1,
\ee
where
$
(\Delta \o)_n =  \o_{n+1}-\o_n
$
is the frequency shift between two neighboring modes.
It follows from (\ref{BFI-incr-max}) that
\be \label{inc-n1}
 \o_{n+1}=\o_n+ \o_n A_n k_n,
\ee
and combination of $A_{n+1}=\sqrt{p} A_n$ and (\ref{inc-n1}) yields
\be \label{1}
\boxed{ A(\o_n+ \o_n A_n k_n)  = \sqrt{p} A_n}
\ee
is called \emph{\textbf{chain equation}}.
\end{frame}

\begin{frame}{}
\begin{center}
\textbf{Computing amplitudes $A=A(\o)$ and energy spectrum}
\end{center}
\bea \label{Taylor}
\sqrt{p} A_n= A(\o_n+ \o_n A_n k_n)= \sum_{s=0} ^ {\infty } \frac {A_n^{(s)}}{s!} \, (\o_n A_n k_n)^{s} \nonumber \\
 = A_n + A_n^{'}\o_n A_n k_n+ \frac 12 A_n^{''}(\o_n A_n k_n)^2+...
\eea
Taking two first terms from RHS of (\ref{Taylor}) and combining with (\ref{1}) we get
\bea \label{2terms}
\sqrt{p} A_n = A_n + A_n^{'}\o_n A_n k_n
\RA (A_n)^{'} = \frac{\sqrt{p}-1}{\o_n k_n } \RA  \nonumber \\
\boxed{A(\o) = (\sqrt{p}-1) \int \frac{d \o}{\o k}+\mbox{const}}
\eea
Substitution of a specific dispersion relation into (\ref{2terms}) gives dependence $A=A(\o)$. Energy spectrum is $E \sim A^2.$ \alert{$\Lleftarrow$ historical moment ;)}

\end{frame}
\begin{frame}{}
\begin{center}
\textbf{Termination and direction  of a cascade}
\end{center}
\emph{Condition for a cascade's termination:}
\be
(\Delta \o)_N= \o_N \frac{(1-\sqrt{p})}{2} + \o_N k_N (A_0 -\frac{(1-\sqrt{p})}{2})=0.
\ee

\emph{Condition for a direct cascade:}

\be
\o_{n+1}-\o_n =\o_n \frac{(1-\sqrt{p})}{2} + \o_n k_n (A_0 -\frac{(1-\sqrt{p})}{2}) >0
\ee

\emph{Condition for an inverse cascade:}

\be
\o_{n+1}-\o_n  <0
\ee

\end{frame}

\begin{frame}{}
\begin{center}
\textbf{Surface water waves, I }\\(K+Shugan, submitted; Shugan+K, in preparation)
\end{center}
Dispersion relation $\o^2=k$ yields ODE
\bea \label{2terms}
\o_n^3 A_n^{'}A_n + (1-\sqrt{p})A_n= 0 \RA
 (A_n)^{'} = \frac{\sqrt{p}-1}{\o_n^3 } \RA  \nonumber \\
 A(\o_n) = (\sqrt{p}-1) \int \frac{d \o_n}{\o_n^3}  \RA
 \boxed{A_n = \frac{(1-\sqrt{p})}{2} (\o_n^{-2}-\o_0^{-2})+A_0}
\eea
and for energy spectrum $E_n \sim A_n^2$ we get
\be
E \sim \o^{-\a}, \quad \mbox{with}  \quad 2\le \a \le 4,
\ee
\emph{\textbf{depending on the details of initial conditions}}.
\end{frame}
\begin{frame}{}
\begin{center}
\textbf{Surface water waves, II}
\end{center}
Condition for a cascade termination in this case turns into ($\o_0=1$)
\be \label{cas-termination}
\o_N^{2}=
\frac{1-\sqrt{p}}{1-\sqrt{p}-2A_0}\,,
\ee
and specific initial conditions
\be \label{in-class}
\frac{(1-\sqrt{p})}{2} =A_0
\ee
yield infinite cascade (\textbf{\emph{transition to the continuous spectrum}}). In this case
\be \label{3}
 A_n \sim \o_n^{-2} \RA A^2_n \sim \o_n^{-4}
\ee
which is \alert{classical Phillips spectrum} for surface water waves.

\end{frame}

\begin{frame}{}
\begin{center}
\textbf{General scheme for computing a cascade} \\(K, in preparation)
\end{center}
\begin{itemize}
\item{} Relation between neighboring amplitudes : \be \label{gen1}
A_{n+1}=\sqrt{p} A_n
\ee
\item{} Maximal increment (\alert{changeable}): \be \label{gen2}
I_{\mathbf{max},n}=\frac{|\o_{n+1}- \o_n|}{\o_n A_n k_n}= 1,
\ee
\item{} Chain equation (\alert{changeable}):\be \label{gen3}
 A(\o_n+ \o_n A_n k_n)  = \sqrt{p} A(\o_n)
\ee
\item{} ODE on amplitude (\alert{changeable}): \be \label{gen4} \sqrt{p} A_n = A_n + A_n^{'}\o_n A_n k_n \ee
\item{} Amplitudes and energy spectrum (\alert{changeable}): \be \label{gen5} A(\o) = (\sqrt{p}-1) \int \frac{d \o}{\o k}+\mbox{const}, \quad E \sim A^2 \                         \ee
    \end{itemize}
\end{frame}

\begin{frame}{}
\begin{figure}
\vskip -0.4cm
\begin{center}
\includegraphics[width=5cm,height=4cm]{KZ-classical.eps}
\includegraphics[width=6cm,height=4cm]{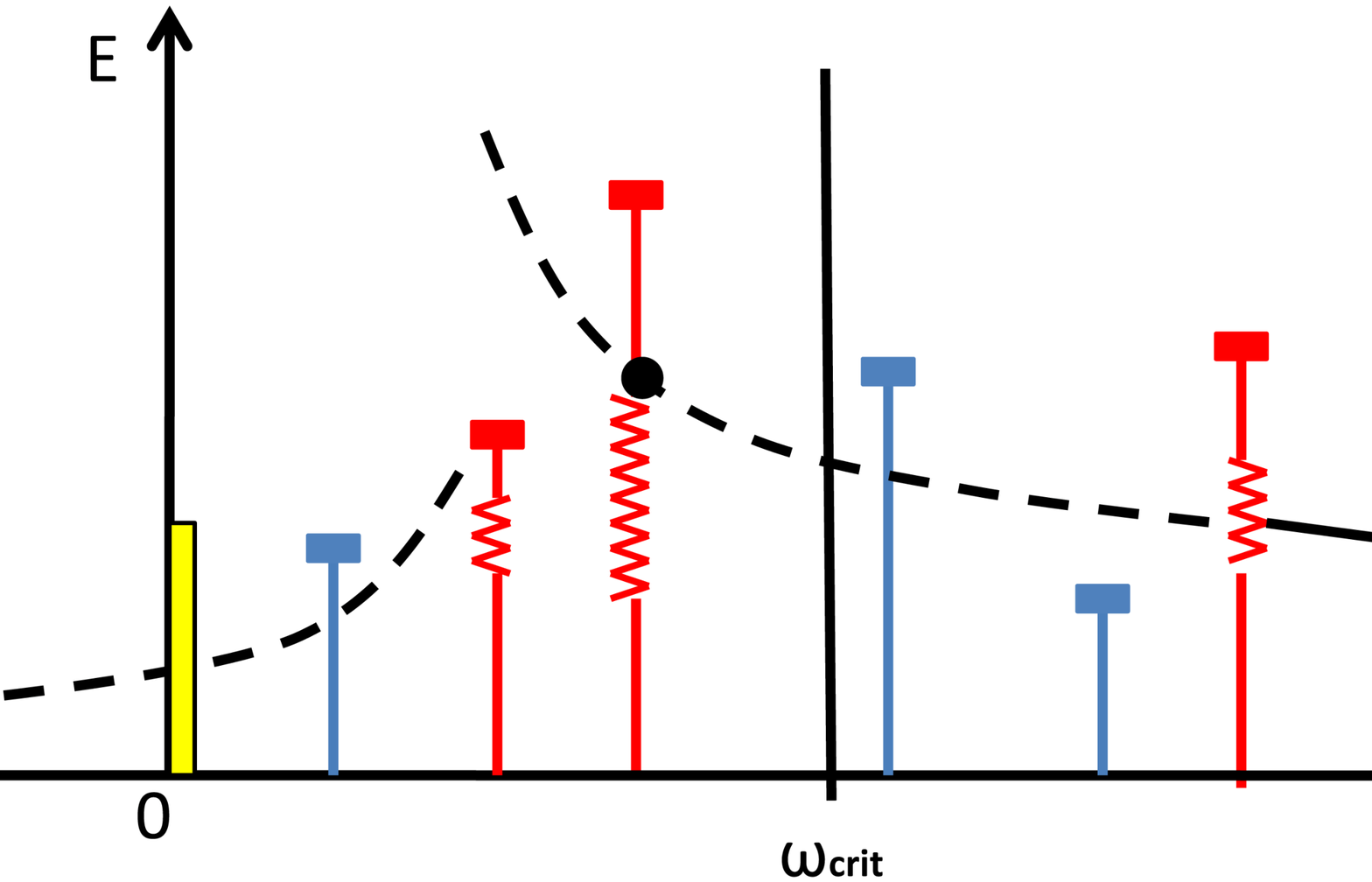}
\end{center}
\end{figure}
\vskip -0.3cm
\begin{itemize}
\item \emph{\textbf{Red vertical solid T-shaped lines having a string-like part}}: resonant and near-resonant modes.
\item \emph{\textbf{Blue vertical solid T-shaped lines}}: fluxless ("frozen") modes.
\item \emph{\textbf{Black dashed curves}}: discrete cascades. \emph{\textbf{Solid part of a curve}} shows continuous "tail" of a cascade.
    \item \emph{\textbf{Vertical yellow rectangular}}: zero-frequency sideband.
\item \emph{\textbf{Black circle}}: source of possible intermittency (chaotic or FPU-like recurrence).
\item \emph{\textbf{Vertical black bold line at $\o_{crit}$}}: "physical" termination of a cascade is due to \alert{INCREASED NONLINEARITY}, not dissipation! "Tails" appearing after this line \alert{are not KZ-spectra!}

\end{itemize}

\end{frame}

\begin{frame}{}
\begin{center}
\textbf{Conclusions: a toy model allows \alert{CONSTRUCTIVELY} to}
\end{center}
\begin{itemize}
\item{} include so-called "discrete effects" in the form resonance clusters and their dynamical systems;
\item{} compute an energy spectrum depending on the form of dispersion relation $\o=\o(k)$ and \alert{initial conditions};
\item{} write out conditions for 1) a cascade's termination, and b) the formation of a direct and/or an inverse cascade;
\item{} explain formation of the narrow zero-band frequency with non-zero energy;
\item{} obtain KZ-energy spectra as a result of some specific initial conditions.
\end{itemize}

\alert{\textbf{The model predicts that direct cascade terminates not due to dissipation but due to the growth of nonlinearity.}}
\end{frame}
\end{document}